\documentclass[referee]{aastex}
\usepackage{emulateapj5}
\slugcomment{Astrophysical Journal Letters, in press (scheduled Dec.~1, 2002)}

\def\spose#1{\hbox to 0pt{#1\hss}}\def\lta{\mathrel{\spose{\lower 3pt\hbox{$\mathchar"218$}}
     \raise 2.0pt\hbox{$\mathchar"13C$}}}
\def\gta{\mathrel{\spose{\lower 3pt\hbox{$\mathchar"218$}}
     \raise 2.0pt\hbox{$\mathchar"13E$}}}
\def\kms{\,{\rm km\,s}^{-1}}
\def\etal{et al.\ }

\def\i{\relax\ifmmode{\rm i}\else\char16\fi}

\def\deg{^\circ}             
\def\frac#1#2{{\textstyle{#1\over#2}}}

\def\dddot#1{\ddot#1\kern-1.4pt\dot{\phantom{#1}}\kern-3pt}

\def\spose#1{\hbox to 0pt{#1\hss}}

\def\=#1{\overline{#1}}

\def\lta{\mathrel{\spose{\lower 3pt\hbox{$\mathchar"218$}}
     \raise 2.0pt\hbox{$\mathchar"13C$}}}
\def\gta{\mathrel{\spose{\lower 3pt\hbox{$\mathchar"218$}}
     \raise 2.0pt\hbox{$\mathchar"13E$}}}

\def\kms{{\rm\,km\,s^{-1}}}
\def\kpc{{\rm\,kpc}}
\def\Mpc{{\rm\,Mpc}}

\def\msun{{\rm\,M_\odot}}

\def\pc{{\rm\,pc}}
\def\cm{{\rm\,cm}}
\def\yr{{\rm\,yr}}

\def\Myr{{\rm\,Myr}}
\def\myr{{\rm\,Myr}}

\def\erg{{\rm\,erg}}

\def\ergpscm{{\rm\,erg\,s}^{-1}{\cm}^{-2}}
\def\K{{\rm\,K}}

\def\annrev #1 #2 {ARA\&A, #1, #2}
\def\aa #1 #2 {A\&A, #1, #2}
\def\aasupp #1 #2 {A\&AS, #1, #2}
\def\aj #1 #2 {AJ, #1, #2}
\def\apj #1 #2 {ApJ, #1, #2}
\def\apjlett #1 #2 {ApJ, #1, #2}
\def\apjsupp #1 #2 {ApJS, #1, #2}
\def\ban #1 #2 {Bull.\ Astron.\ Inst.\ Netherlands, #1, #2}
\def\mn #1 #2 {MNRAS, #1, #2}
\def\nature #1 #2 {Nat, #1, #2}
\def\pasj #1 #2 {PASJ, #1, #2}
\def\pasp #1 #2 {PASP, #1, #2}

\shorttitle{Isolated HII Region in Virgo Cluster}
\shortauthors{Gerhard et al.}

\begin{document}
\title{Isolated Star Formation: \\
       A Compact HII Region in the Virgo Cluster\footnotemark}
\author{Ortwin Gerhard$^1$, Magda Arnaboldi$^{2,3}$,}
\author{Kenneth C.~Freeman$^4$, Sadanori Okamura$^5$} 
\affil{$^1$Astronomisches Institut, Universit\"at Basel, Venusstrasse
7, CH-4102 Binningen, Switzerland \\
$^2$I.N.A.F., Osservatorio Astronomico di Capodimonte, 80131 Naples, Italy \\
$^3$I.N.A.F., Osservatorio Astronomico di Pino Torinese, 10025
Pino Torinese, Italy \\
$^4$R.S.A.A., Mt.~Stromlo Observatory, 2611 ACT, Australia \\
$^5$Department of Astronomy, School of Science, University of Tokyo,
Tokyo, 113-0033, Japan}

\begin{abstract}
  We report on the discovery of an isolated, compact HII region in the
  Virgo cluster. The object is located in the diffuse outer halo of
  NGC 4388, or could possibly be in intracluster space.  Star
  formation can thus take place far outside the main star forming
  regions of galaxies.  This object is powered by a small starburst
  with an estimated mass of $\sim 400\msun$ and age of $\sim 3\myr$.
  From a total sample of $17$ HII region candidates, the present rate
  of isolated star formation estimated in our Virgo field is small,
  $\sim 10^{-6} \msun {\rm arcmin}^{-2} \yr^{-1}$. However, this mode
  of star formation might have been more important at higher redshifts
  and be responsible for a fraction of the observed intracluster stars
  and total cluster metal production.  This object is relevant also
  for distance determinations with the planetary nebula luminosity
  function from emission line surveys, for high-velocity clouds and
  the in situ origin of B stars in the Galactic halo, and for local
  enrichment of the intracluster gas by Type II supernovae.
\end{abstract}


\footnotetext[1]{Based on observations carried out at UT4 of the VLT,
  Paranal, Chile, which is operated by the European Southern
  Observatory.}

\section{Introduction}

Stars are usually observed to form in galaxies, that is, in disks, dwarfs,
and starbursts. In radio galaxies, star formation may be triggered
by energetic jet outflows (e.g., Bicknell \etal 2000). The HII region
we describe here shows that isolated star formation takes place in the
diffuse outskirts of galaxies, at the boundary of, if not already in,
Virgo intracluster space.

In nearby galaxy clusters, a diffuse intracluster star component has
been inferred from surface brightness measurements (Bernstein \etal
1995) and detection of individual stars (e.g., Arnaboldi \etal 1996,
Ferguson \etal 1998, Feldmeier 2002).  Its origin may be explained
readily by dynamical processes acting on low-surface brightness disks
and dwarfs, which unbind the stars from these galaxies (Moore \etal
1999).  However, gas may be efficiently removed from infalling
galaxies by ram pressure stripping (Quilis \etal 2000, Gavazzi \etal
2001), or may be tidally dissolved, or could fall into the cluster as
pristine clouds. Some of this gas may form stars, which would also
contribute to the diffuse component. We discuss here the current rate
of such star formation from isolated compact HII regions (ICHIIs) and
from jet induced star formation in a Virgo cluster field.

\section{An HII region in Virgo: observed properties}
The target object was found in an emission line survey for planetary
nebulae in a Virgo intracluster field, centred at $\alpha(J2000)
12:25:31.9,\, \delta(J2000) 12:43:47.7$, using H$\alpha$ and [OIII]
narrow band and V+R broad band filter photometry (Arnaboldi \etal
2002, Okamura \etal 2002), with Suprime-Cam on the Subaru Telescope.
Because of the large H$\alpha$ to [OIII] flux ratio the (unresolved)
object was classified as a candidate compact HII region.

A spectrum was taken at UT4 of the VLT at Paranal, on the night of
April 14, 2002, with FORS2 in MOS mode.  The observations were carried
out with GRISM-150I and the order separation filter GG435+81, giving a
wavelength coverage of $4500 - 10200$ \AA\ and a dispersion of $6.7$
\AA\ pix$^{-1}$.  The slit width was $1.4$ arcsec, and
the angular scale along the slitlet was $0.126$ arcsec pix$^{-1}$.  The
total exposure time was $7\times1800s$.  The nights were clear but not
photometric; the mean seeing was better than 1.0 arcsec.
Spectrophotometric standard stars were observed at the beginning and
end of the night, but cirrus clouds at these times made the flux calibration
uncertain.

\begin{figure*}
\begin{center}
\plotone{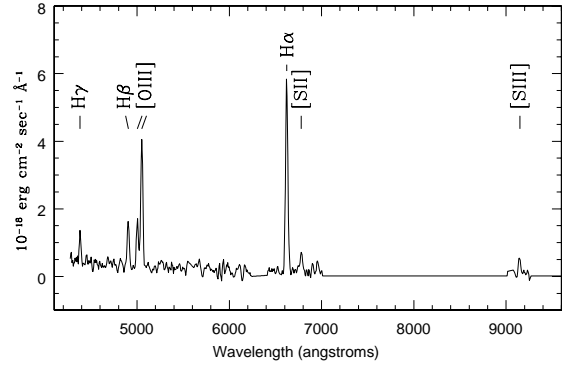}
\caption[dsb]{\small Observed emission spectrum.}
\label{spectrum}
\end{center}
\end{figure*}

The data reduction was carried out using standard tasks in IRAF, using
an arc lamp wavelength calibration and observations of a
spectrophotometric standard star.  The spectrum was corrected for
atmospheric extinction, using a table for La Silla\footnote{No
  correction for atmospheric refraction is needed since FORS2 has an
  atmospheric dispersion corrector.}.

\def\llam{\lambda\lambda}

The wavelength and flux-calibrated spectrum is shown in Figure 1:
it clearly shows a blue
continuum and a number of emission lines at Virgo redshift: H$\alpha$,
H$\beta$, H$\gamma$, [OIII]$\lambda 5007$ and $\lambda 4959$,
[SII]$\llam6717+6731$, and [SIII]$\lambda 9069$. 
We do not resolve H$\alpha$ and [NII]$\lambda 6548$, 
and the [SII]$\llam
6717,6731$ lines, respectively, and we do not see the weaker
[OI]$\lambda6300$, [OIII]$\lambda 4363$, and [SIII]$\lambda6312$
lines. The H$\gamma$ line lies close to the blue end of the spectrum,
and the [SIII]$\lambda 9069$ line in a region of strong sky emission.

\begin{table*}
\caption[Line fluxes.]
{Observed and reddening-corrected emission line fluxes relative to H$\beta$.
}
\label{tablines}
\vspace{0.5cm} 
{\setlength\tabcolsep{0.45em}
\centerline{%
}}
\begin{tabular}{lccc}
\tableline
&&&\\
Line   & $\lambda$ &  observed flux        & de-reddened flux \\
       &  \AA      &  rel.\ to H$\beta$    & rel.\ to H$\beta$ \\
\tableline 
&&&\\
H$\gamma$         &  4340     &  0.58   & 0.71  \\
H$\beta$          &  4861     &  1.00   & 1.00   \\
{[OIII]}          &  4959     &  1.03   & 0.99  \\
{[OIII]}          &  5007     &  3.05   & 2.91   \\
{[NI]}            &  5200     &  0.08   &      \\
H$\alpha$         &  6563     &  4.30   & 2.89   \\
{[NII]}           &  6583     &  0.75   & 0.50  \\
{[SII]}           & 6717+6731 &  0.40   & 0.26  \\
{[SIII]}          &  9069     &  0.30   & 0.14  \\
&&&\\
\tableline
&&&\\
\multicolumn{4}{l}{[NI] identification uncertain.} \\ 
\multicolumn{4}{l}{[NII]$\lambda 6583$ flux from two-Gaussian fit.} \\
\multicolumn{4}{l}{H$\alpha$ flux is corrected for corresponding 
[NII]$\lambda 6548$ emission.} \\
\end{tabular}
\end{table*}

The observed emission lines and their uncorrected and
reddening-corrected fluxes normalized to H$\beta$ are listed in 
Table 1. The errors in the fluxes are approximately 10\% of the 
H$\beta$ flux, and larger at the blue edge of the spectral range.
The H$\alpha$ and [NII]$\lambda 6548$ lines are unresolved, with the
[NII]$\lambda 6583$ line appearing in the red wing of H$\alpha$.  From
a two-Gaussian fit to the combined emission we estimate the line ratio
[NII]$\lambda 6583 /({\rm H} \alpha + {\rm [NII]}\lambda 6548) \simeq
0.17$.  This is consistent with the value expected for an HII region
with the large observed [OIII]$\lambda 5007$/H$\beta$ ratio; see
below.  The H$\alpha$ flux given in the table is that measured as the
total flux in the line reduced by one third that in [NII]$\lambda
6583$. The observed Balmer decrement H$\alpha$/H$\beta$ then becomes
4.3.

If the intrinsic Balmer decrement has the theoretical value of 2.89
for large optical depth (case B) and temperature $T=10^4\K$, then
E(B-V)=0.40 and A$_{\rm V}$=1.23 mag.  Most of this reddening is
intrinsic to the source, as the galactic absorption to nearby galaxies
in the Virgo cluster is about A$_{\rm V}\simeq 0.12$.  Line fluxes
were corrected for reddening using this value of E(B-V) and the
reddening curve of Cardelli \etal (1989).  Compared to H$\beta$, the
corrected H$\gamma$ flux is about 1.5 times larger than expected;
however, we cannot regard the H$\gamma$ flux as reliable, because the
line lies right at the edge of the spectrum where uncertainties in the
grism response, sky subtraction and determination of the continuum
level will be at their worst.

The corrected line ratios [OIII]$\lambda 5007$/H$\beta=2.91$,
[NII]$\lambda 6583$/H$\alpha = 0.17$, and [SII] $\llam
6717,6731$/H$\alpha = 0.09$ place the object clearly into the domain
of HII regions in Figs.~1,2 of Veilleux \& Osterbrock (1987; VO87).
This is also confirmed by the weakness (non-detection) of
[OI]$\lambda 6300$ in the spectrum.

Because the flux calibration using our standard stars is uncertain, we
transformed it to the Jacoby (1989) m$_{\rm OIII}$ calibration of the
imaging photometry from Suprime-Cam (Arnaboldi \etal 2002). This gives
a total m$_{\rm OIII}=25.7$ magnitude for this object.  Because of the
[OIII] filter width used in the emission line survey, this flux
includes both the [OIII]$\llam 4959,5007$ lines, as well as the flux
from the continuum.  Using the spectrum to measure the fraction of
light in the continuum, we can correct to the magnitude in the
$\lambda 5007$ line only: m$_{5007}=26.15$. From the standard
star--calibrated spectrum, we would get m$_{5007}=-2.5\log F_{5007}
-13.74=25.1$. The difference gives us a factor 2.63, by which to
rescale $F_{5007}$ to match the photometric calibration. The rescaled
total flux in the [OIII] $\lambda 5007$ \AA\ line is $F_{5007}=
11.0\times 10^{-17} \ergpscm$.  With E(B-V)=0.4 the dereddened flux
becomes $F_{5007}=4.0\times 10^{-16} \ergpscm$.  The fluxes in the
other emission lines can then be inferred from the dereddened line
ratios in Table 1; in particular the H$\alpha$ flux becomes $3.9\times
10^{-16} \ergpscm$.

The measured V-band continuum flux from the spectrum, when applying
the same calibration and correcting for reddening, becomes
$F_V=7.0\times 10^{-16} \ergpscm$. The corresponding dereddened
apparent magnitude is m$_V=24.2$. Absolute magnitudes and fluxes will
be computed for a nominal Virgo distance modulus of 31.16,
corresponding to $17\Mpc$ distance (Tonry \etal 2001). However, we
will sometimes keep the distance dependence by writing $D=d_{17}\Mpc$.
The continuum $m_V=24.2$ then gives $M_V=-7.0$.

\section{Physical parameters}

{\bf Metallicity:} The high values of [OIII]/H$\alpha=1.0$ and
[OIII]/H$\beta=2.9$ indicate subsolar metallicity. The VO87 diagrams are
not ideal for estimating metallicities, as discussed by Dopita \etal
(2000), but from Figs.~2,3 of that paper we estimate $Z\simeq 0.4$ and an
ionization parameter $q\simeq 4 \times 10^7$ ($\log U=-2.9$).  
Both very low and high
metallicities are not consistent with their model results and these
line ratios. A more accurate determination is in principle possible
using the $S_{23}$ parameter, defined as
$S_{23}=$([SII]$\llam6717,6731$+[SIII]$\llam9069,9532$) / H$\beta =
0.75$, where we have used the theoretical ratio
[SIII]$\lambda9532$/[SIII]$\lambda9069=2.48$. From the empirical
calibration of D\'iaz \& P\'erez-Montero (2000) we then find
$12+\log({\rm O/H})\simeq 8.08\pm0.2$ (about 0.15 to 0.25 solar with
or without the depletion factor used by Dopita \etal 2000). In our
case, $S_{23}$ has additional uncertainties due to the weak [SII] line
and the bright NIR sky lines, which might affect the [SIII] flux. In
the following, we therefore adopt $Z\simeq 0.4$.

{\bf Temperature, density, ionization parameter:} Unfortunately, the
S/N of the spectrum is not large enough to detect the weak
[OIII]$\lambda 4363$ and [NII]$\lambda 5755$ lines used to determine
electron temperature, nor do we have lines to determine the electron
density.  In the following we use $T_e=10^4\K$ when needed.

Estimates of the ionization parameter from the Sulfur lines, using
equations (8) and (6) of D\'iaz \etal (2000) give similar values to
that from the VO87 diagrams:
[SII]$\llam6717,6731$/[SIII]$\llam9069,9532 \simeq 0.53$ gives $\log
U = -2.5$, and [SII]$\llam6717,6731$/H$\beta \simeq 0.26$ and
metallicity 0.4 solar gives $\log U = -2.9$.

{\bf Luminosities:} From above, the total V-band luminosity is
\begin{equation}
 L_V = 4\pi D^2 F_V = 2.4\times 10^{37} d_{17}^2 \erg\, {\rm s}^{-1}. 
\end{equation}
The H$\alpha$ flux is
\begin{equation}
 L_{{\rm H}\alpha}= 4\pi D^2 F_{{\rm H}\alpha} = 1.3 \times 10^{37} 
                                         d_{17}^2 \erg\, {\rm s}^{-1}. 
\end{equation}
The total number of H-ionizing photons (Osterbrock 1989) with
recombination coefficients $\alpha_B$ and 
$\alpha_{{\rm H}\alpha}^{\rm eff}$ for $T=10^4\K$ is
\begin{equation}
 Q({\rm H}^0)=2.96\, L_{{\rm H}\alpha} /h\nu_{{\rm H}\alpha}
     = 1.3 \times 10^{49} d_{17}^2\, {\rm s}^{-1}. 
\end{equation}

{\bf Stellar mass and age of starburst:} The ratio $Q({\rm H}^0)/L_V$
decreases rapidly with the mass of the most massive surviving O stars,
i.e., the age of the starburst. Using the Starburst99 model of
Leitherer \etal (1999) for metallicity 0.4 solar and normal Salpeter
IMF, we determine an age of $ 3.3 \Myr$. Once this is known, the Lyman
continuum luminosity can be used to infer the total mass in the
corresponding starburst; we obtain $ 400 \msun$. The number of 
O stars involved is 1--2, so these numbers must be considered as
averages.

{\bf Mass and size of gas cloud:} We can estimate the total mass
of ionized hydrogen from (Osterbrock 1989)
\begin{equation}
 M_{\rm HII}= Q({\rm H}^0) m_p/[n_e\alpha_B]
       =420 n_{100}^{-1} d_{17}^2 \msun ,
\end{equation}
where $n_{100}=n_e/100\cm^{-3}$ and $m_p$ denote the electron density
and proton mass. Around early O stars, the H Str\"omgren radius
is
\begin{equation}
 r_{\rm HII} = 
  \left[ { M_{\rm HII} (1+y^+) \over (4\pi/3)n_e m_p} \right]^{1/3}
  \simeq 3.5 n_{100}^{-2/3} d_{17}^{2/3} \pc,
\end{equation}
where $y^+\simeq0.1$ is the fraction by number of ionized helium.
This is indeed much smaller than the photometric spatial
resolution (about $0''.7=57 d_{17} \pc$). The column density is
\begin{equation}
 N_{\rm HII}=M_{\rm HII}/(\pi r_{\rm HII}^2) = 1.3\times 10^{21} 
                        n_{100}^{1/3} d_{17}^{2/3} \cm^{-2}.
\end{equation}

From the inferred extinction and metallicity we may estimate
the intervening hydrogen column density from the local interstellar medium
relation (Bohlin, Savage \& Drake 1978), using 
\begin{equation}
 N({\rm H}) \simeq 5.9 \times
10^{21} \cm^{-2} {\rm mag}^{-1} {\rm E}_{B-V} Z_\odot/Z \sim 6 \times
10^{21} \cm^{-2}.
\end{equation}
If the neutral hydrogen is in a much denser shell than the HII, then we 
can use the HII radius also to estimate its total mass
\begin{equation}
 M_{\rm H}=\pi r_{\rm HII}^2 \times 2 N({\rm H}) m_p
            = 3600 \msun.
\end{equation}
The star formation efficiency would then be around 10\%.

\section{Discussion}

The compact HII-region is located about $3.'4$ (17 kpc projected
distance) north and $0.'9$ (4.4 kpc) west of NGC 4388, almost
perpendicular to the disk plane of this galaxy and $45\deg$ away
from the nearest part of the very extended emission-line region
(VEELR) discussed by Yoshida \etal (2002).  The radial velocity
inferred from the emission lines is $2670\kms$, whereas the galaxy has
$v_r=2520\kms$. The near-coincidence of these numbers may indicate
that the HII region is, or once was, part of the NGC 4388 system. From
the large radial velocity relative to the Virgo center ($v_r \sim
1.8\sigma_{\rm Virgo}$) and from the Tully-Fisher distance of NGC 4388
(Yasuda \etal 1997), both are probably falling through the cluster
core.

This HII region is powered by a small stellar association or star
cluster, with an estimated mass of $\sim 400\msun$ and age of $\sim 3
\myr$ for a normal IMF. If it is a young cluster, it must have a
radius smaller than that of the HII region ($\sim 3.5\pc$). Clusters
with these parameters have short relaxation times and dissolve by
internal dynamical processes; within a few $10^8\yr$ the stars would
be added to the diffuse stellar population nearby.

The formation of such objects is thus possible far outside the main
star forming regions in galaxies.  Perhaps the most plausible
explanation for the observed position and velocity of the HII region
relative to NGC 4388 is that it is already unbound and moving on a
different orbit in the cluster potential. In this case we would be
seeing intracluster star formation.  If on the other hand its true
distance from NGC 4388 is comparable to the projected distance so that
it is still bound, we would have discovered a small star-forming knot
in the far outer halo of NGC 4388 far from any other star formation
activity. This would have a velocity in the frame of the galaxy
comparable to the circular velocity ($\simeq 200\kms$) of NGC 4388. In
fact, in the Subaru field that contained our ICHII region, Arnaboldi
\etal (2002) found a small sample of similar candidate objects, among
which 7 candidates are located in the outskirts of M86, at a distance
of 10--15 kpc in a disturbed region which probably also contains
diffuse H$\alpha$ emission, and 1 such object in the outer parts of
M84.

What has triggered the recent onset of star formation in this remote 
cloud?
Notice that with its radial velocity, it will have moved $\sim 4 \kpc$
relative to the Virgo cluster, and $\sim 500\pc$ relative to NGC 4388,
in the lifetime of the massive stars.  Thus these stars will
essentially have formed in situ. A possible model could be that the
cloud was compressed after entering the hot intracluster medium (ICM).
While the typical pressure of the ICM ($n_e T \sim 10^4 \K/\cm^3$) is
far smaller than the internal pressure inferred from typical HII
region parameters($n_e\sim 100 / \cm^3$; $T \sim 10^4 \K$), it could
be comparable to that of the surrounding neutral or molecular cloud
traced by the absorption.  The ionized region may be reexpanding into
the surrounding cloud and into the ICM -- the mass of the cluster is
not large enough to bind the HII region.

Reanalysing the Subaru field photometry, we have also detected a
number of extended emitters in both [OIII] and H$\alpha$, with colors
similar to the confirmed compact HII region. These are peaks in more
extended emissions, and lie at distances 11--33 kpc along the
direction of the VEELR identified
by Yoshida \etal (2002) towards the NE of NGC 4388. These objects have
relatively low excitation ([OIII]/H$\alpha\simeq 0.6$, in agreement
with Fig.~6 of Yoshida \etal 2002), and show continuum emission near
the limiting magnitude of the combined (V+R) image (Arnaboldi et al.
2002).  Their half-light radii are similar in the combined continuum
and in the [OIII] and H$\alpha$ images, which further supports the
notion that these distant sources have underlying continuum emission,
and are thus ionized by OB stars rather than by the nucleus of NGC
4388. From the discussion of Yoshida \etal (2002) it is likely that
this star formation is induced by the jet or ionization cone.

Yoshida \etal (2002) proposed that the VEELR observed to the NE
of NGC 4388 could be the debris of a past interaction. Although
our ICHII region is $2.6$ arcmin away from the nearest part of
the VEELR, it could also be associated with such tidal debris.
The gas in this ICHII region seems unlikely to come from ram
pressure stripping of the gas in NGC 4388, because the peculiar
velocity of the ICHII region relative to the systemic velocity
of the Virgo cluster is even larger than for NGC 4388 itself.
(We thank B.~Moore for this comment.)

The total intracluster star formation rate (ISFR) estimated from the
candidates in this field is small, however. Counting the 8
ICHII candidates near M86 and M84 plus 3 intracluster candidates,
including that studied here, and the 6 best extended candidates, gives
a total of 17 star-forming HII regions in the Subaru field. If the
density of these objects is typical for the Virgo cluster core, these
would correspond to $\sim10^3$ such objects throughout Virgo.  If we
adopt similar parameters to those inferred for the HII region here
(conservatively), we obtain from these candidates a total ISFR
of $\sim 10$ times $\sim 400 \msun$ per $3\myr$, in a surveyed area of
918 arcmin$^2$, i.e., $\sim 10^{-6} \msun {\rm arcmin}^{-2} \yr^{-1}$.
For comparison, the intracluster luminosity inferred from planetary
nebulae in the Subaru field is $\sim 10^7 {\rm L}_{B,\odot} {\rm
  arcmin}^{-2}$. However, it appears not impossible that at higher
redshifts, when the environment of infalling galaxies was more
gas-rich, formation of stars directly from this intracluster gas could
have been an important part of the origin of intracluster stars in
Virgo.

The existence of compact HII regions in Virgo, beyond its intrinsic
interest, is relevant for a few other issues as well. Firstly, the
massive stars ionizing the gas explode as type II supernovae and
enrich the Virgo ICM with metals. This process adds to the main metal
content of the ICM, which is believed to have occured at high redshift
when most of the stars in the cluster ellipticals and bulges formed
(Renzini 1999). From the ISFR inferred above, the present expected
supernova rate is $\sim 10^{-8} {\rm arcmin}^{-2} \yr^{-1}$. SN II
from isolated star formation could provide the enrichment inferred for
the Ly$\alpha$ clouds in Virgo (Tripp \etal 2002). Because our ICHII
region contains only 1-2 O stars, observations of the enrichment from
such objects in isolated regions might give constraints on the element
yields of individual supernovae. Newly formed stars from this material
could then have a range of abundance ratios similar to old Galactic
halo stars (Argast \etal 2000).

Secondly, compact HII regions may affect distance determinations
with the planetary nebula luminosity function (PNLF, Jacoby \etal
1992).  The [OIII] luminosity of this ICHII region places it at the
bright end of the luminosity function inferred from [OIII] emission
line surveys.  In a sufficiently deep offband control image its
continuum light would be visible, leading to removal from the PN
sample. However, this does require a deep image, and moreover there
are a few HII candidates in the Subaru field for which the continuum
is not detected.  A few unrecognized such objects per galaxy or
intracluster field could cause the distance to these fields to be
underestimated. It is unclear whether this may account for the lower
average distance inferred by the PNLF compared to surface brightness
fluctuations ( Ciardullo \etal 2002).

Thirdly, if compact ICHII regions exist in galaxies generally, they
could be the birth places of distant B stars in the Galactic halo,
some of which are too far from the disk to have been ejected from
there into the halo (Conlon \etal 1992). Most likely, their birth
places would be distant high velocity clouds. At this time, however,
there is no evidence for star formation in galactic high-velocity
clouds.

\acknowledgements{}

We thank R.~Scarpa for efficient help at UT4, J.~Alcal\'a for
independently checking line fluxes, and A.~Capetti for a helpful
discussion. OG thanks the Swiss Nationalfonds for support under grant
20-64856.01. This research has made use of the NASA extragalactic
database.

\end{document}